\documentclass[reprint,amsmath,amssymb,prx,superscriptaddress]{revtex4-1}
\usepackage{graphicx}
\usepackage{bm}
\usepackage{booktabs}
\usepackage[capitalize]{cleveref}
\usepackage{siunitx}

\begin{document}
\title{A Nonlinear Enhanced Microresonator Gyroscope}

\author{Jonathan~M.~Silver}
 \email{jonathan.silver@npl.co.uk}
\affiliation{National Physical Laboratory, Hampton Road, Teddington TW11 0LW, UK}
\affiliation{City, University of London, Northampton Square, London EC1V 0HB, UK}
\author{Leonardo~Del~Bino}
\author{Michael~T.~M.~Woodley}
\affiliation{National Physical Laboratory, Hampton Road, Teddington TW11 0LW, UK}
\affiliation{Heriot-Watt University, Edinburgh EH14 4AS, UK}
\author{George~N.~Ghalanos}
\author{Andreas~\O.~Svela}
\author{Niall~Moroney}
\affiliation{National Physical Laboratory, Hampton Road, Teddington TW11 0LW, UK}
\affiliation{Blackett Laboratory, Imperial College London, London SW7 2AZ, UK}
\author{Shuangyou~Zhang}
\affiliation{National Physical Laboratory, Hampton Road, Teddington TW11 0LW, UK}
\author{Kenneth~T.~V.~Grattan}
\affiliation{City, University of London, Northampton Square, London EC1V 0HB, UK}
\author{Pascal~Del'Haye}
\affiliation{National Physical Laboratory, Hampton Road, Teddington TW11 0LW, UK}
\affiliation{Max Planck Institute for the Science of Light, 91058 Erlangen, Germany}
\date{\today}

\begin{abstract}
We demonstrate a novel optical microresonator gyroscope whose responsivity to rotation is enhanced by a factor of around $10^4$ by operating close to the critical point of a spontaneous symmetry breaking transition between counterpropagating light. We present a proof-of-principle rotation measurement using a resonator with a diameter of \SI{3}{mm}. In addition, we characterise the dynamical response of the system to a sinusoidally varying rotation, and show this to be well described by a simple theoretical model. We observe the universal critical behaviors of responsivity enhancement and critical slowing down, both of which are beneficial in an optical gyroscope.
\end{abstract}

\maketitle

\section{Introduction}

The ultrahigh quality factors achievable with optical microresonators offer the possibility of realising a novel form of optical gyroscope~\cite{DellOlio2014, Ma2017} with a fraction of the size, weight, power consumption, and cost of fiber-optic gyroscopes (FOGs) and ring laser gyroscopes (RLGs)~\cite{DellOlio2014}. Advances in microfabrication techniques, combined with innovative methods of measuring the Sagnac effect~\cite{Post1967} in microresonators, are beginning to make mass-producible, chip-based optical gyroscopes a real possibility. Examples of such methods include phase modulation schemes~\cite{ma2015double, wang2015resonator}, the Pound-Drever-Hall (PDH) technique~\cite{liang2017resonant}, and stimulated Brillouin scattering or lasing~\cite{Li2017, lai2019observation}. Several techniques for enhancing the rotation sensitivity have been proposed and demonstrated, including phase difference traversal~\cite{Wang2016}, dual-resonator reciprocal measurement~\cite{Khial2018}, tunable dispersion~\cite{Zhang2016a}, and the use of exceptional points in non-Hermitian systems~\cite{Ren2017, Sunada2017, lai2019observation}. This last example has attracted considerable interest as the response of the resonator is proportional to the square root of the rotation velocity at an exceptional point, potentially leading to a sensitivity increase of several orders of magnitude.

One of the key features of ultrahigh-Q optical microresonators is the strong Kerr nonlinearity that they exhibit at modest input powers of milliwatts or even microwatts, leading to important effects such as frequency comb generation~\cite{Del'Haye2007, Stern2018, zhang2019sub}. Recently, Kerr interaction between counterpropagating light waves in a bidirectionally-pumped microresonator was found to give rise to spontaneous symmetry breaking~\cite{del2017symmetry, Cao2017, Woodley2018}. This occurs because the Kerr interaction between the counterpropagating waves, a form of cross-phase modulation, is stronger than self-phase modulation by a factor of 2 in a dielectric solid, which means that differences between the counterpropagating circulating powers self-amplify via resonance frequency splittings and consequent pump-resonance detuning differences. As well as enabling novel nonreciprocal optical components~\cite{DelBino2018} and optical memories~\cite{Daniel2012}, this symmetry breaking drastically enhances the response of the resonator to rotation~\cite{Kaplan1981, Wang2014a} and near-field perturbations~\cite{Wang2015,svela2019spontaneous}. This occurs when the system is operating close to the critical point of the symmetry breaking transition, with a cube-root response, and hence divergent reponsivity to small rotation velocities, at the critical point itself. Divergent sensitivity to external perturbations is actually a universal feature of critical points in general~\cite{stanley1971phase}, occuring in systems as diverse as the Higgs mechanism~\cite{higgs1964broken}, ferromagnetism, liquid-gas critical points, superconductivity~\cite{PhysRev.108.1175} and superfluidity~\cite{landau1941theory}. Another such universal feature is critical slowing down~\cite{stanley1971phase}, where the characteristic timescale of the system's response to a perturbation diverges towards the critical point. This manifests itself in our system as though the rotation velocity signal were being acted on by a low-pass filter that behaves like an integrator above its cut-off frequency, which approaches zero towards the critical point~\cite{silver2019criticalPaperArxiv}.

Here we report on an enhanced gyroscope based on this principle in a silica microrod resonator~\cite{DelHaye2013apl} with diameter \SI{2.8}{mm} and $Q=2.9\times10^8$ coupled to a tapered optical fiber, using laser light at \SI{1550}{nm}. We directly observe both responsitivity enhancement and critical slowing down. Both of these effects are beneficial for a gyroscope, as together they cause the system to integrate rotation velocity to yield rotation angle when it is operating sufficiently close to the critical point.

\section{Theoretical Background}
In an optical ring resonator of diameter $D$ and refractive index $n_0$ rotating in its plane at angular velocity $\Omega$, the Sagnac effect causes the resonance frequencies of counterpropagating pairs of modes at vacuum wavelength $\lambda$ to differ by an amount~\cite{Post1967}
\begin{equation}
\Delta\omega = 2\pi\frac{D\Omega}{n_0\lambda}.\label{eq:rawSagnac}
\end{equation}
For \SI{1550}{nm} light in a silica resonator with a diameter of a few millimetres, rotating at \SI{1}{deg/s}, this is just tens of Hz, which is already 3 to 4 orders of magnitude smaller than the resonator's linewidth, even for a state-of-the-art Q factor of $10^9$~\cite{Lee2012}. Thus, to turn such a resonator into a gyroscope that significantly improves on microelectromechanical systems (MEMS) devices~\cite{Passaro2017}, it is necessary to be able to detect splittings of $<10^{-5}$ of the linewidth, which is extremely challenging with a direct measurement such as via the PDH technique.

\begin{figure}
\includegraphics*[width=0.95\columnwidth]{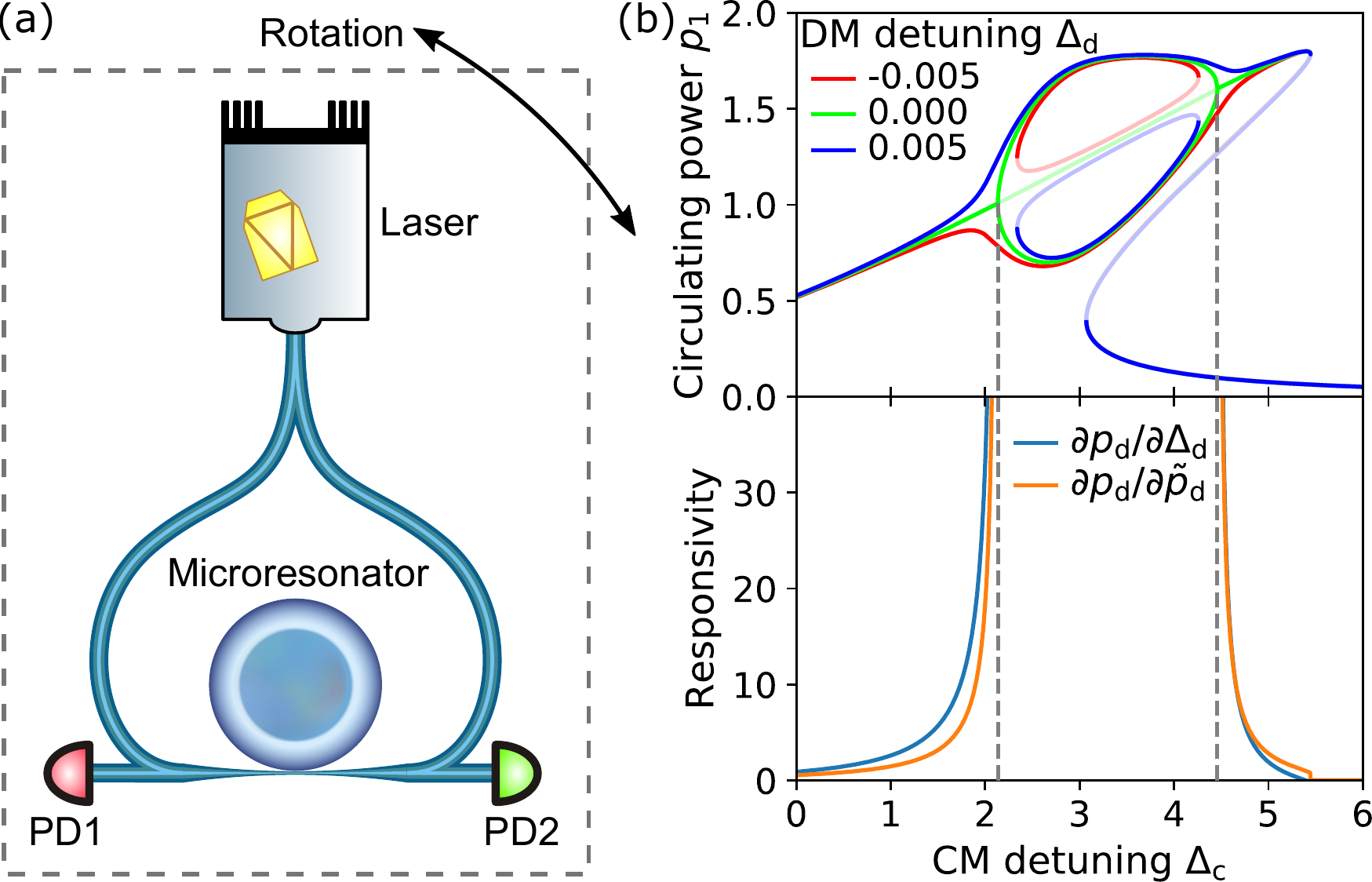}
\caption{(a) Simplified schematic of the nonlinear enhanced microresonator gyroscope. A high-Q optical ring resonator is pumped equally in both directions with narrow-linewidth continuous-wave laser light via a tapered optical fiber. The power and detuning are set so as to place the resonator near the critical point of the symmetry breaking regime \cite{del2017symmetry, Woodley2018} and the transmitted powers, indicative of the circulating powers, are detected on two photodiodes (PD). Rotation in the plane of the resonator causes a difference between the readings on PD1 and PD2. (b) Top: circulating power $p_1$ in one of the two directions vs.\ common-mode (CM) detuning $\Delta_\mathrm{c}$ for three values of the differential-mode (DM) detuning $\Delta_\mathrm{d}$ and equal pump powers $\tilde{p}_{1,2}=1.8$ (see \cref{tab:dlqs}), showing the sensitivity enhancement near the critical points (of which we use the left one). All quantities are in their dimensionless forms (see \cref{tab:dlqs}). Faint lines correspond to unstable solutions. By symmetry, the curves for $p_2$ are identical but with $\Delta_\text{d}$ negated, so when $p_1$ takes the upper blue curve, $p_2$ takes the lower red one etc. Bottom: Partial derivatives of the DM circulating power $p_\mathrm{d}$ with respect to the DM detuning $\Delta_\mathrm{d}$ and pump power $\tilde{p}_\mathrm{d}$ for the case $\Delta_\mathrm{d}=0$ (see \cref{eq:dpddDd,eq:dpddptd}).}\label{fig:sch}
\end{figure}

Here we investigate how the sensitivity enhancement~\cite{Kaplan1981, Wang2014a} that exists near the critical point of Kerr-mediated symmetry breaking between counterpropagating light fields~\cite{del2017symmetry, Woodley2018} may be used to overcome this problem. The basic form of the setup is illustrated in \cref{fig:sch}(a). Monochromatic light of the same power and frequency is coupled evanescently into a high-Q ring resonator in both the clockwise and counterclockwise directions. The pump power and detuning are chosen so as to place the resonator close to the critical point of the symmetry breaking.

\begin{table}
\caption{Dimensionless quantities used in this manuscript. The subscripts 1 and 2 refer to the two counterpropagating directions, and `c' and `d' to common- and differential-mode (CM and DM) combinations of these. $\eta_\text{in}$ is the resonant in-coupling efficiency equal to $4\kappa\gamma_0/\gamma^2$ where $\kappa$, $\gamma_0$ and $\gamma=\gamma_0+\kappa$ are the coupling, intrinsic and total half-linewidths respectively. $P_{\text{in,}1,2}$ and $P_{\text{circ,}1,2}$ are the pump and circulating powers respectively. $P_0 = \pi n_0^2V/(n_2\lambda QQ_0)$ is the characteristic in-coupled power required for Kerr nonlinear effects, where $n_0$ and $n_2$ are the linear and nonlinear refractive indices, $V$ is the mode volume, and $Q=\omega_0/(2\gamma)$ and $Q_0=\omega_0/(2\gamma_0)$ are the loaded and intrinsic quality factors respectively for cavity resonance frequency $\omega_0$ (without Kerr shift). $\mathcal{F}_0 = \Delta\omega_\text{FSR}/(2\gamma_0)$ is the cavity's intrinsic finesse for free spectral range $\Delta\omega_\text{FSR}$, $\omega_{1,2}$ are the two pump frequencies and $\Delta$ is the value of $\Delta_1=\Delta_2$ at the critical point.}
\label{tab:dlqs}
  \begin{center}
    \begin{tabular}{cll}
      
      \toprule
      \textbf{Symbol} & \textbf{Description} & \textbf{Formula}\\
      \midrule
     $\tilde{p}_{1,2}$ & Pump powers & $\eta_\text{in}P_{\text{in,}1,2}/P_0$\\[5pt]
     $p_{1,2}$ & Circulating powers & $2\pi P_{\text{circ,}1,2}/(\mathcal{F}_0 P_0)$\\[5pt]
      $\Delta_{1,2}$ & \parbox{3.4cm}{\raggedright Pump detunings from resonance frequency without Kerr shift} & $(\omega_0-\omega_{1,2})/\gamma$\\[17pt]
      $\delta_{1,2}$ & \parbox{3.4cm}{\raggedright Pump detuning offsets from critical point} & $\Delta_{1,2}-\Delta$\\[10pt]
      $\tilde{e}_{1,2}$ & Pump field amplitudes & $\tilde{p}_{1,2} = \left|\tilde{e}_{1,2}\right|^2$\\[5pt]
      $e_{1,2}$ & \parbox{3.4cm}{\raggedright Circulating field amplitudes} & $p_{1,2} = \left|e_{1,2}\right|^2$\\[12pt]
      $\epsilon_{1,2}$ & \parbox{3.4cm}{\raggedright Fractional pump power perturbations} & $\tilde{p}_{1,2} = \tilde{p}(1+\epsilon_{1,2})$\\[10pt] \parbox{1.7cm}{$\tilde{p}_\text{c,d}$, $p_\text{c,d}$,\\$\Delta_\text{c,d}$, $\delta_\text{c,d}$,\\$\epsilon_\text{c,d}$} & \parbox{3.4cm}{\raggedright CM and DM components \strut} & \parbox{3.08cm}{\raggedright For $X\!\in\!\{\tilde{p},p,\Delta,\delta,\epsilon\},$ \\ $X_\text{c} = (X_1\!+\!X_2)/2,$ \\ $X_\text{d} = (X_1\!-\!X_2)/2$} \\
      \bottomrule
    \end{tabular}
  \end{center}
\end{table}

Throughout this text, we shall express quantities in the dimensionless forms listed in \cref{tab:dlqs}. Time is normalised by the inverse half-linewidth $1/\gamma$, and angular frequencies by $\gamma$. Since the dynamics we are interested in take place over timescales much longer than $1/\gamma$, cavity ringdown effects are negligible, meaning that the values of $p_{1,2}$ are directly related to the transmitted powers detected on the photodiodes in the counterpropagating directions (see \cref{fig:sch}).

In the steady state, $p_{1,2}$ obey the following pair of simultaneous equations~\cite{silver2019criticalPaperArxiv}:
\begin{equation}
p_{1,2} = \frac{\tilde{p}_{1,2}}{1+(p_{1,2}+2p_{2,1}-\Delta_{1,2})^2}.\label{eq:steadystate}
\end{equation}
Note the factor of 2 in front of the counterpropagating circulating power, which represents the ratio between the strengths of cross- and self-phase modulation. For symmetric pump powers $\tilde{p}_{1,2} = \tilde{p}$ and detunings $\Delta_{1,2}=\Delta$, the symmetric solution $p_{1,2} = p$ thus satisfies
\begin{equation}
p = \frac{\tilde{p}}{1+\left(3p-\Delta\right)^2}.\label{eq:ptpRelat}
\end{equation}
For $\tilde{p}$ above the threshold $8/(3\sqrt{3})\simeq1.54$, a symmetry-broken regime exists in which the symmetric solution is unstable and is replaced by two stable asymmetric solutions that map to each other under swapping of the two directions. The critical points satisfy the condition~\cite{silver2019criticalPaperArxiv}
\begin{equation}
\left(p-\Delta\right)\left(3p-\Delta\right)=-1.\label{eq:critPtCond}
\end{equation}
This symmetry breaking is illustrated in \cref{fig:sch}(b) for $\tilde{p}=1.8$. The upper panel shows how a detuning difference of just 1\% of the half-linewidth causes the two circulating powers to differ by a large proportion near the critical points. The responsivity in fact diverges as the critical point is approached, as shown in the lower panel, which makes the system useful as a gyroscope since rotation is directly linked to $\Delta_\text{d}$. However, the system is simultaneously divergently responsive to differences in pump power, which means that the sensitivity to rotation is limited by the stability of the pump power difference. These responsivities, for $\tilde{p}_\text{d}=p_\text{d}=\Delta_\text{d}=0$, are given by:
\begin{align}
\frac{\partial p_\text{d}}{\partial\Delta_\text{d}} &= \frac{1}{1+(p_\text{c}-\Delta_\text{c})(3p_\text{c}-\Delta_\text{c})}\label{eq:dpddDd}\\
\frac{\partial p_\text{d}}{\partial\tilde{p}_\text{d}} &= \frac{2p_\text{c}(3p_\text{c}-\Delta_\text{c})}{1+(p_\text{c}-\Delta_\text{c})(3p_\text{c}-\Delta_\text{c})}.\label{eq:dpddptd}
\end{align}
Note that $(3p_\text{c}-\Delta_\text{c})>0$, i.e.\ the laser must be on the blue side of the Kerr-shifted resonance, for symmetry breaking to be observed~\cite{Woodley2018}. Furthermore, the region where the denominator (and hence both derivatives) is negative corresponds to the unstable symmetric solution between the two critical points, shown as a faint green line in \cref{fig:sch}(b).

A full analysis~\cite{silver2019criticalPaperArxiv} shows that very close to the critical point, the dynamics are governed to leading order by the equation
\begin{equation}
\dot{y}=-y^3+xy+z\label{eq:critDynEq}
\end{equation}
where
\begin{align}
x&=\frac{5p-2\Delta}{4}\delta_\text{c}+\frac{2\Delta-3p}{4}p\epsilon_\text{c}\nonumber\\
y&=\;\sqrt{\!\frac{15p^2-4p\Delta-4}{8p\left(3p-\Delta\right)}}\,p_\text{d}\nonumber\\
z&=\sqrt{\frac{p\left(3p-\Delta\right)\left(15p^2-4p\Delta-4\right)}{8}}\left(\delta_\text{d}+p\epsilon_\text{d}\right)\label{eq:critDynScal}
\end{align}
in which $\epsilon_\text{c}$ and $\epsilon_\text{d}$ are CM and DM fractional pump power purturbations respectively, as detailed in \cref{tab:dlqs}.

\section{Experimental Methods}\label{sec:expmet}
\begin{figure}
\includegraphics*[width=0.95\columnwidth]{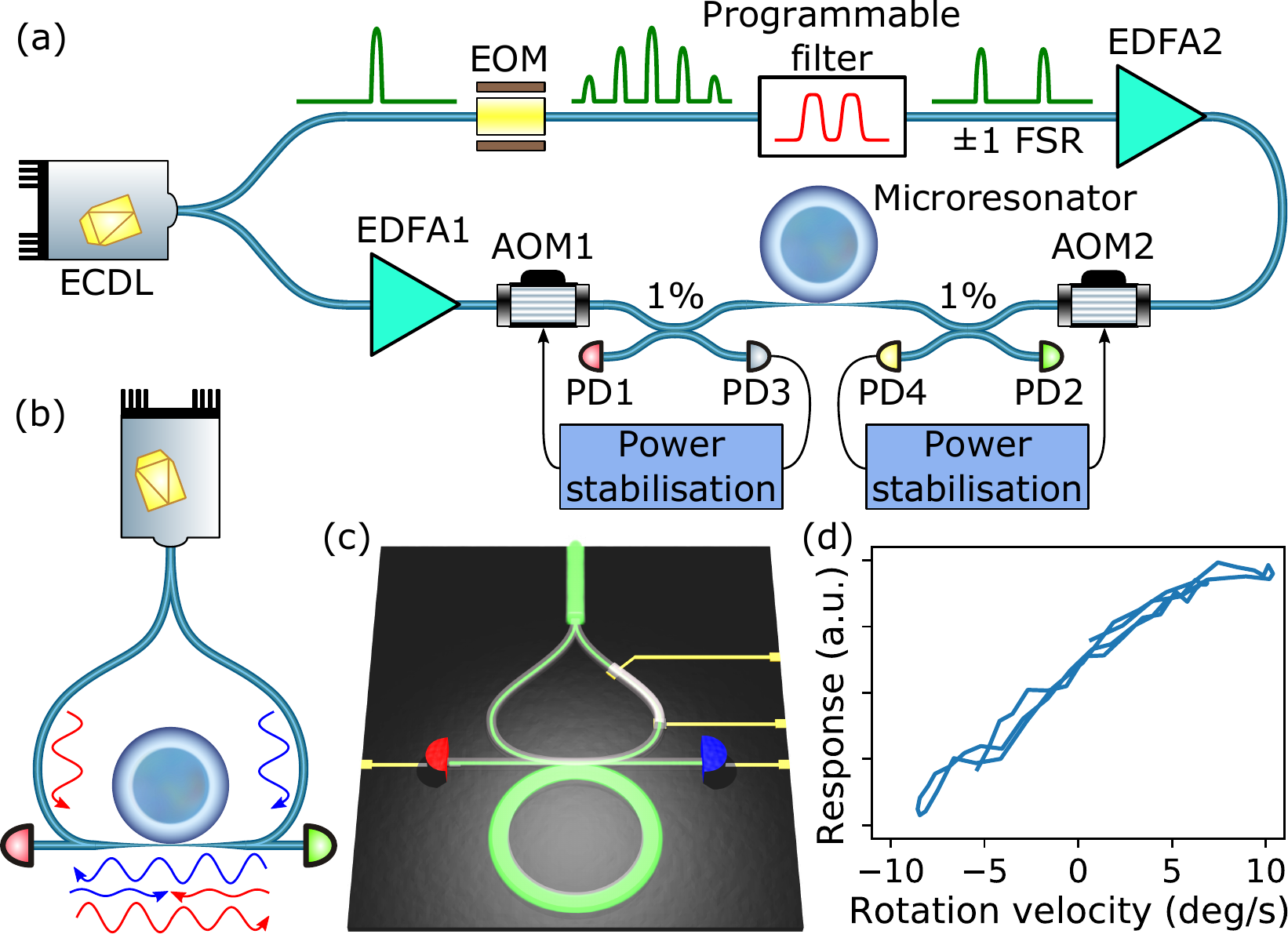}
\caption{(a) Optical circuit of the experiment. ECDL = external cavity diode laser, EDFA = erbium-doped fiber amplifier, AOM = acousto-optic modulator, PD = photodiode, FSR = free spectral range (of the microresonator, around \SI{23.7}{GHz}), 1\% = 1\% directional coupler. (b) Illustration of the interference effect that prevents us from using the same pump frequency in both directions. (c) Illustration of a possible monolithic chip-based realization of the optical circuit that would not suffer from this interference problem. (d) System response (difference between readings on PD1 and PD2) vs.\ rotation velocity measured using a MEMS gyroscope, indicating a sensitivity of around \SI{2}{deg/s}.}\label{fig:set}
\end{figure}
The optical circuit used in the experiment is summarized in \cref{fig:set}(a). Light from a narrow-linewidth tunable external-cavity diode laser (ECDL) at around \SI{1550}{nm} is amplified before being coupled into the microresonator via a tapered optical fiber. Due to the fiber-based nature of the setup, pumping the resonator with the same frequency of light in both directions would have led to a prohibitive problem with interference, depicted in \cref{fig:set}(b). Unavoidable spurious back-reflections from fiber components, connections, etc.\ would interfere with counterpropagating light, and acoustic and thermal noise in the fibers would cause the relative phase of the interfering waves, and hence the pump powers seen by the resonator, to fluctuate. Even if the back-reflections are at the level of \SI{-40}{dB}, as is typical in our experiment, the pump powers would vary by a few percent, which in our system would limit the rotation sensitivity to around one revolution per second.

To solve this, it was necessary to use different pump frequencies for each direction. Detuning the two pumps by a fraction of the resonator's linewidth, and compensating this with a power difference to get back to the critical point~\cite{Woodley2018} would have helped somewhat as it would wash out the interference phase at long timescales. However, the resonator still reacts to the percent-level oscillating pump power difference at short timescales, making it impossible for it to stay on average very close to the critical point and exhibit the enormous responsivity enhancement associated therewith. To solve this, the two pump frequencies were offset by the resonator's free spectral range (FSR), so that the counterpropagating waves would continue to have near-perfect spatial overlap within the resonator (due to being coupled into the same mode family) and hence experience the same Kerr interaction, whilst the interference would be removed completely. This dramatically improves the pump power stabilities, allowing the system to remain much closer to the critical point. However, offsetting the counterpropagating pump frequencies by one FSR makes the system sensitive to temperature-related drifts in the FSR as these now translate into pump detuning differences; this in turn was solved by making one of the pumps an equal composition of light one FSR higher and one FSR lower than the other. As shown in \cref{fig:set}(a), this was achieved by sending the light through an electro-optic modulator (EOM) driven with an RF frequency equal to the FSR, followed by a programmable wavelength filter. This allowed the system to be placed at a point where it is first-order-insensitive to variations in the FSR, which was achieved by fine-tuning the EOM's RF drive frequency until the transmitted powers measured on PD1 and PD2 were locally stationary with respect thereto. 

It is worth noting that splitting one pump into two separate frequencies increases its effective self-phase modulation strength by a factor of 3/2, since each frequency experiences a combination of its own self-phase modulation and the twice-as-strong cross-phase modulation from the other frequency. A full quantitative treatment of this situation is presented in~\cite{silver2019criticalPaperArxiv}, which shows how the critical point can be attained by compensating this imbalance through pump power and/or detuning differences. However, this does not significantly change the critical dynamics, so that \cref{eq:steadystate} still holds.

It is important to stress that these somewhat convoluted measures for overcoming the interference problem are likely not to be needed if the optical circuit is realised on a monolithic chip-based platform (illustrated in \cref{fig:set}(c)), as is the aim for this gyroscope. This is because the very short optical paths and monolithic nature of the setup would ensure that the phase of any interference remains extremely stable, meaning that both pumps can have the same frequency without their powers fluctuating significantly.

Turning again to our setup (\cref{fig:set}(a)), the pump powers were monitored using PD3 and PD4, and continuously stabilised by feeding back to the RF powers driving AOM1 and AOM2. The RF frequencies to the AOMs were used to control the pump detuning difference between the two directions in order to tune the system to the critical point. The transmitted powers were recorded via PD1 and PD2. Rotation of the resonator was achieved by mounting the entire optical circuit except for the laser and EDFA1, plus some of the electronics, on a structure suspended from the ceiling that acted as a torsional pendulum. Light was transmitted to this part of the setup via polarization-maintaining (PM) fibers and linear polarizers to ensure that the incoming polarizations remained constant as the setup was rotated. Although most fibers on the rotating setup itself were non-PM, care was taken to minimize the effect of polarization drift on the pump powers seen by the (polarization-sensitive) resonator. This was done by filtering the polarization just before the pick-offs to PD3 and PD4, by minimizing lengths of fiber, and by securing any loose sections of fiber. Fiber polarization controllers were placed immediately before each polarization-sensitive element of the circuit. A third pick-off (1\% directional coupler) was placed just after the programmable filter to allow the spectrum to be monitored on an optical spectrum analyser. Around \SI{50}{mW} of optical power was sent into the tapered fiber in each direction.

The angular velocity of the rotating setup was detected using a chip-based MEMS gyroscope mounted rigidly to it. This was used to produce \cref{fig:set}(d), in which the difference between the transmitted powers measured on PD1 and PD2 -- the ``response'' -- was recorded alongside the MEMS gyroscope reading as the torsional pendulum setup rotated freely back and forth a few times over the course of 42 seconds. This measurement indicates a rotation sensitivity of around \SI{2}{deg/s}, which is limited by the accuracy with which the pump powers seen by the resonator are able to be stabilized due to various sources of noise in the setup. The resonator's enhancement factor $\partial p_\text{d}/\partial\Delta_\text{d}$ in this measurement was around $10^4$.

To characterise the dynamical response of the system, it was necessary to perform extended measurements under rapidly varying angular velocity. Since the torsional pendulum was hand-actuated and had limited allowable angular acceleration, rather than physically rotating the setup in these measurements, the Sagnac splitting was simulated by changing the pump detuning difference $\Delta_\text{d}$ via the AOM driving frequencies. In fact, since in the region very close to the critical point the system is much more sensitive to $\Delta_\text{d}$ than to $\Delta_\text{c}$, this effect could be achieved by varying just one of the AOM frequencies; although this would cause both the differential- and common-mode detunings $\Delta_\text{d}$ and $\Delta_\text{c}$ to change, the effect of the variation in $\Delta_\text{c}$ was negligible. After adjusting the laser frequency and detuning difference to reach the critical point, waiting a few minutes for the resonator to thermalize, optimizing all the polarizations, and tuning the EOM frequency to the aforementioned stationary point, the RF frequency to AOM1 was modulated sinusoidally at \SI{500}{Hz} and a lock-in measurement of the system's response to this was made. This measurement was performed for a range of both $\Delta_\text{c}$ (accessed via the laser frequency) and the DC offset of $\Delta_\text{d}$ (accessed via the DC offset to the driving frequency of AOM1), over the course of several hours. To compensate for any thermal drift in the FSR that would take the system away from the stationary point, the EOM frequency was automatically adjusted back to this point at regular intervals. This was achieved by modulating it at \SI{10}{kHz}, performing an in-phase lock-in measurement of the system's response, and stabilizing this to zero by feeding back to the EOM frequency's DC offset, before turning off the modulation and retaining the offset's last value.

The readings on PD1 and PD2 were converted to the dimensionless quantities $p_{1,2}$ by the following procedure: firstly, quantities proportional to $p_{1,2}$ and which shall be referred to as ``coupled powers'', were calculated as the differences between the photodiode readings and their ``baseline values'' taken when the pump was completely out of resonance. Next, the value of $p/\tilde{p}$ at the critical point was calculated as the mean of the two ratios $p_{1,2}/\tilde{p}_{1,2}$ between the coupled powers at the critical point and at maximum coupling (when the pump is perfectly on resonance and $p_{1,2}=\tilde{p}_{1,2}$). This was then used to find $p$ (along with $\Delta$) at the critical point by combining \cref{eq:ptpRelat,eq:critPtCond}, thus giving the constants of proportionality between the coupled powers and $p_{1,2}$.

Values of $\delta_\text{d}=\Delta_\text{d}$ were found by dividing the detuning difference offset from the critical point by twice the cavity half-linewidth $\gamma$, which was measured using a cavity ringdown technique~\cite{Lecaplain2016} to be around \mbox{$2\pi\times3$\SI{30}{kHz}}. Furthermore time and angular frequency of modulation are normalised by $1/\gamma$ and $\gamma$, respectively. Importantly, the thermal shift of the resonance frequencies in silica resonators is huge~\cite{Carmon2004}, at almost two orders of magnitude larger than the Kerr shift. This meant that $\Delta_\text{c}$, or its offset $\delta_\text{c}$ from the critical point, could not be obtained directly from the laser frequency. Instead, they were calculated from $p_\text{c}$ and the critical point values $p$ and $\Delta$ via the relation $\partial p_\text{c}/\partial \Delta_\text{c} = 1/4$ that holds at the critical point.

\section{Results and Discussion}

\begin{figure}
\includegraphics*[width=0.98\columnwidth]{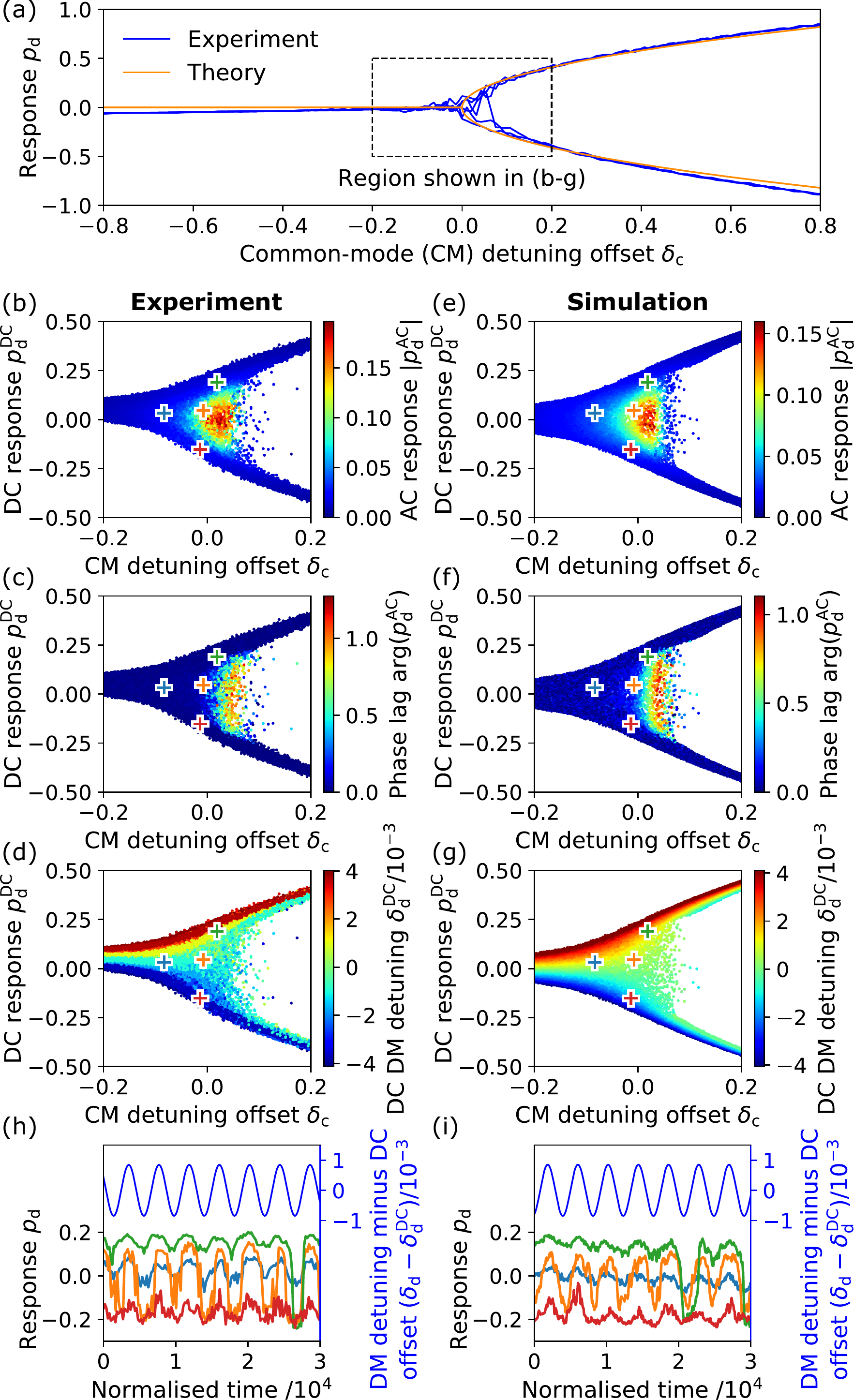}
\caption{Experimental results (a--d, h) with theoretical curves (a) and simulations (e--g, i) based on our simplified critical dynamics model with no free parameters. All quantities are given in their dimensionless forms (see \cref{tab:dlqs} and \cref{eq:modu,eq:demodu}). (a) Differential-mode (DM) circulating power $p_\mathrm{d}$ vs.\ common-mode (CM) detuning offset from critical point $\delta_\mathrm{c}$ for six sweeps of the laser frequency, showing the symmetry breaking in either direction, with theoretical curves overlaid. (b--g) Magnitude (b, e) and phase (c, f) of the demodulated AC response $p_\mathrm{d}^\mathrm{AC}$ to a sinusoidally modulated rotation at \SI{500}{Hz} ($1.5\times10^{-3}$ in dimensionless units) vs.\ CM detuning offset $\delta_\mathrm{c}$ and DC response $p_\mathrm{d}^\mathrm{DC}$ for a range of DM detuning DC offsets $\delta_\mathrm{d}^\mathrm{DC}$ (d, g). Rather than physically rotating the setup, the detuning difference between the pumps was modulated via the RF frequency to AOM1 (see Fig.\ 1) to mimic the Sagnac splitting. The (half peak-to-peak) modulation amplitude was \SI{560}{Hz}, or $\delta_\mathrm{d}^\mathrm{AC}=8.4\times10^{-4}$ in dimensionless units, corresponding to \SI{23}{deg/s} of rotation velocity. (h, i) System response $p_\mathrm{d}$ vs.\ time (normalised by the inverse half-linewidth $1/\gamma$) for four different sets of parameters $\delta_\mathrm{c}$ and $\delta_\mathrm{d}^\mathrm{DC}$ indicated by correspondingly-colored crosses in (b--g), alongside the input modulation signal $\delta_\mathrm{d}-\delta_\mathrm{d}^\mathrm{DC}$.}\label{fig:res}
\end{figure}

\Cref{fig:res}(a) shows the differential-mode circulating power $p_\text{d}$ vs.\ the common-mode detuning offset $\delta_\text{c}$ from the critical point, for six downward sweeps of the laser frequency across the critical point. Note that $p_\text{d}$ randomly chooses one sign or the other during each sweep as the detuning passes the critical point. These are overlaid with the stable steady-state solutions to \cref{eq:critDynEq} for $z=0$, namely $y=0$ for $x<0$ and $y=\pm\sqrt{x}$ for $x>0$. The scaling factors from $x$ and $y$ to $\delta_\text{c}$ and $p_\text{d}$ are calculated from the experimentally determined values of $p=1.02$ and $\Delta=1.85$ at the critical point via \cref{eq:critDynScal}. The fact that the curvature of the parabola matches the data without fitting provides validation for the model. The dashed rectangle in \cref{fig:res}(a) indicates the region shown in (b--g). 

\Cref{fig:res}(b--d) depict the system's measured response to a sinusoidally modulated DM detuning offset 
\begin{equation}
\delta_\text{d}=\delta_\text{d}^\text{DC}+\delta_\text{d}^\text{AC}\cos(\Omega_\text{mod}t),\label{eq:modu}
\end{equation}
where $\delta_\text{d}^\text{AC}$ and $\Omega_\text{mod}$ are $8.4\times10^{-4}$ and $1.5\times10^{-3}$ respectively in dimensionless units. Each data point represents a measurement lasting \SI{100}{ms}, or \mbox{50 periods} of the modulation, from which both the time averages of both the DC and demodulated DM circulating power,
\begin{equation}
p_\text{d}^\text{DC} = \left<p_\text{d}\right>\quad\text{and}\quad p_\text{d}^\text{AC} = 2\left<p_\text{d}e^{i\Omega_\text{mod}t}\right>\label{eq:demodu}
\end{equation}
respectively, were obtained. Throughout the measurement period, the laser frequency was scanned with a \SI{0.1}{Hz} triangle wave over a range of more than \SI{100}{MHz} (due to the large thermal nonlinearity of silica discussed at the end of Section~\ref{sec:expmet}~\cite{Carmon2004}), to access a range of common-mode detunings $\delta_\text{c}$. The value of $\delta_\text{c}$ for each data point was derived from the time-averaged measured value of $p_\text{c}$ as discussed at the end of Section~\ref{sec:expmet}. After every 1000 data points, the value of the DC DM detuning $\delta_\text{d}^\text{DC}$ was changed, and was cycled in ths way through 70 values evenly spaced in the range $\pm4\times10^{-3}$.

The magnitude and phase of $p_\text{d}^\text{AC}$, as well as $\delta_\text{d}^\text{DC}$, are respresented by the colors of the data points in \mbox{\cref{fig:res}(b--d)} respectively, in all of which the horizontal and vertical coordinates are $\delta_\text{c}$ and $p_\text{d}^\text{DC}$ respectively. The reason why $p_\text{d}^\text{DC}$ rather than the independent variable $\delta_\text{d}^\text{DC}$ is used as the vertical axis is because there is noise on the pump power difference that has a very similar effect to noise on $\delta_\text{d}^\text{DC}$. Even though this noise is at the level of $10^{-4}$, it is significant in this measurement since the system's responsivities to fractional differences in pump power and to differences in normalised detuning are roughly equal, as can be seen from the expression for $z$ in \cref{eq:critDynScal}. Despite all the steps taken to reduce relative pump power fluctuations in the setup, there were still a few sources of this noise. Plotting the data in this way allows panels (b) and (c) to remain unaffected by this noise. 

From \cref{fig:res}(b) we can see a dramatic increase in the responsivity around the critical point, which is a universal feature of critical dynamics. From \cref{fig:res}(c) we observe a phase lag in the AC response approaching $\pi/2$, which shows that the system is acting more like an integrator than a proportional amplifier. This is a clear indicator of critical slowing down, another universal critical behavior~\cite{stanley1971phase}. Although it is responsible, together with the aforementioned noise, for limiting the maximum AC responsivity, it is also useful as it allows the gyroscope to measure rotation angle rather than simply rotation velocity.

The data for $\delta_\text{d}^\text{DC}$ vs.\ $\delta_\text{c}$ and $p_\text{d}^\text{DC}$ (\cref{fig:res}(d)) were fitted with a theoretical function based on the steady-state solution to \cref{eq:critDynEq}, and the residuals of this fit vs.\ time were then used to find the frequency spectrum of the equivalent noise on $\delta_\text{d}^\text{DC}$. As expected due to the multiple sources of noise, the power spectral density (PSD) has an approximate overall $1/f$ dependency; a fit of this function to the PSD vs.\ frequency (on a log-log plot) gives a coefficient of $1.5\times10^{-9}$ in units of $\delta_\text{d}^2$, which in our setup is equivalent to an rms noise on the measured rotation velocity of around \SI{0.25}{deg/s/\sqrt{Hz}} at \SI{1}{kHz}. The measurement noise in units of $\delta_\text{d}$ is approximately equal to the noise on the DM fractional pump power $\epsilon_\text{d}$, and can be reduced by moving to a monolithic chip-based waveguide circuit as illustrated in \cref{fig:set}(c). On such a platform, the pump power difference and polarizations would be passively stable, rather than relying on separate feedback circuits and polarization filters for each direction. In addition, the orders-of-magnitude-lower path length noise would mean that interference from back-reflections (see \cref{fig:set}(b)) would not lead to significant pump power fluctuations, and thus the same pump frequencies could be used in both directions, greatly simplifying the circuit. The only active element that would be required in the optical circuit would be a variable attenuator for fine-tuning the balance between the two pump powers, which could be based on any of a number of electro-optic or thermo-optic effects. Furthermore, for a given noise level in units of $\delta_\text{d}$, the noise level on the measured rotation velocity can be reduced by increasing the Q factor or diameter of the resonator, as can be seen from \cref{eq:rawSagnac} and the relation $\Delta\omega = 2\gamma\delta_\text{d}$.

\Cref{fig:res}(h) shows some traces of $p_\text{d}$ vs.\ dimensionless time $\gamma t$ for the different values of $\delta_\text{c}$ and $p_\text{d}^\text{DC}$, alongside the sinusoidal modulation curve of $\delta_\text{d}$ with the DC offset removed. The greatly increased gain of the system, both to the sinusoidal modulation of $\delta_\text{d}$ and to the noise on $p_\text{d}$, near the critical point is apparent by comparing the traces. Note also how for the green trace, which occurs just inside the symmetry-broken region, a  relatively large excursion in the noise that coincides with the correct part of the modulation cycle can cause $p_\text{d}$ to switch momentarily from one symmetry-broken state to the other.

\Cref{fig:res}(e--g, i) show the correponding results of a simulation of \cref{eq:critDynEq} with the same parameters as in the experiment, with the scaling factors between the dimensionless experimental variables and $x$, $y$ and $z$ calculated using \cref{eq:critDynScal}. In the simulation, $1/f$ noise was added to the sinusoidally-modulated input variable $z$ with the same coefficient as was obtained from the fit to the experimental noise spectrum. The experimental and simulated plots agree well, with any discrepancy originating from uncertainty in the measured values of parameters such as $\gamma$, as well as from the slight asymmetry of the system due to the two pump frequencies being sent in one of the directions (see Section~\ref{sec:expmet}).

\section{Conclusion and Outlook}

We have demonstrated a proof-of-principle nonlinear enhanced microresonator gyroscope that operates at the critical point of Kerr-induced symmetry breaking between counterpropagating light in a bidirectionally pumped ring resonator. Rotation in the plane of the microresonator causes a tiny `seed' splitting between the counterpropagating resonance frequencies due to the Sagnac effect, which is then amplified by four orders of magnitude via a positive feedback cycle between resonance splittings and circulating power differences. It can then be `read off' from the microresonator as a large difference between the in-coupled powers in the two directions.

In addition to demonstrating rotation measurement with a sensitivity of around \SI{2}{deg/s}, we have characterised the dynamical response of the system to a sinusoidally varying simulated rotation generated by modulating the pump detuning difference. These measurements were shown, via a numerical simulation, to be well described by a simple theoretical model for the system's critical dynamics~\cite{silver2019criticalPaperArxiv}, and provide direct evidence for two universal critical behaviors, namely responsivity enhancement and critical slowing down. 

At the critical point, the system exhibits divergent responsivity not only to rotation, which is equivalent to pump detuning differences between the two directions, but also to pump power differences. This means that in order to achieve a certain rotation sensitivity, it is necessary to stabilise the pump power difference to an equally high degree, since the system responds about equally to a certain fractional pump power difference as it does to a Sagnac splitting that is the same fraction of the linewidth. We believe that by moving to a monolithic chip-based platform with waveguide circuits and resonators, differential pump power stabilities orders of magnitude higher than in this experiment could be achieved passively, and by also increasing the Q factor and/or diameter of the resonator, rotation senitivities approaching those of today's fiber optic or ring laser gyroscopes could be achieved in a simple device with a fraction of the size, weight, power consumption, and cost.

\begin{acknowledgments}
This work was supported by the Royal Academy of Engineering and the Office of the Chief Science Adviser for National Security under the UK Intelligence Community Postdoctoral Fellowship Programme. The authors also acknowledge funding from Horizon 2020 Marie Sklodowska-Curie Actions (MSCA) (Contract No.\ 748519, CoLiDR), a Horizon 2020 Marie Sklodowska-Curie COFUND Action (GA-2015-713694), National Physical Laboratory Strategic Research, Horizon 2020 European Research Council (ERC) (No.\ 756966, CounterLight), and the Engineering and Physical Sciences Research Council (EPSRC). L.\ D.\ B.\ and M.\ T.\ M.\ W. acknowledge funding from EPSRC through the Centre for Doctoral Training in Applied Photonics. A.\ \O.\ S. acknowledges funding from the Aker Scholarship and from EPSRC through the Quantum Systems Engineering programme.
\end{acknowledgments}

\bibliography{Gyro_dynamics_noTheory_v3}

\begin{thebibliography}{35}%
\makeatletter
\providecommand \@ifxundefined [1]{%
 \@ifx{#1\undefined}
}%
\providecommand \@ifnum [1]{%
 \ifnum #1\expandafter \@firstoftwo
 \else \expandafter \@secondoftwo
 \fi
}%
\providecommand \@ifx [1]{%
 \ifx #1\expandafter \@firstoftwo
 \else \expandafter \@secondoftwo
 \fi
}%
\providecommand \natexlab [1]{#1}%
\providecommand \enquote  [1]{``#1''}%
\providecommand \bibnamefont  [1]{#1}%
\providecommand \bibfnamefont [1]{#1}%
\providecommand \citenamefont [1]{#1}%
\providecommand \href@noop [0]{\@secondoftwo}%
\providecommand \href [0]{\begingroup \@sanitize@url \@href}%
\providecommand \@href[1]{\@@startlink{#1}\@@href}%
\providecommand \@@href[1]{\endgroup#1\@@endlink}%
\providecommand \@sanitize@url [0]{\catcode `\\12\catcode `\$12\catcode
  `\&12\catcode `\#12\catcode `\^12\catcode `\_12\catcode `\%12\relax}%
\providecommand \@@startlink[1]{}%
\providecommand \@@endlink[0]{}%
\providecommand \url  [0]{\begingroup\@sanitize@url \@url }%
\providecommand \@url [1]{\endgroup\@href {#1}{\urlprefix }}%
\providecommand \urlprefix  [0]{URL }%
\providecommand \Eprint [0]{\href }%
\providecommand \doibase [0]{https://doi.org/}%
\providecommand \selectlanguage [0]{\@gobble}%
\providecommand \bibinfo  [0]{\@secondoftwo}%
\providecommand \bibfield  [0]{\@secondoftwo}%
\providecommand \translation [1]{[#1]}%
\providecommand \BibitemOpen [0]{}%
\providecommand \bibitemStop [0]{}%
\providecommand \bibitemNoStop [0]{.\EOS\space}%
\providecommand \EOS [0]{\spacefactor3000\relax}%
\providecommand \BibitemShut  [1]{\csname bibitem#1\endcsname}%
\let\auto@bib@innerbib\@empty
\bibitem [{\citenamefont {Dell'Olio}\ \emph {et~al.}(2014)\citenamefont
  {Dell'Olio}, \citenamefont {Tatoli}, \citenamefont {Ciminelli},\ and\
  \citenamefont {Armenise}}]{DellOlio2014}%
  \BibitemOpen
  \bibfield  {author} {\bibinfo {author} {\bibfnamefont {F.}~\bibnamefont
  {Dell'Olio}}, \bibinfo {author} {\bibfnamefont {T.}~\bibnamefont {Tatoli}},
  \bibinfo {author} {\bibfnamefont {C.}~\bibnamefont {Ciminelli}},\ and\
  \bibinfo {author} {\bibfnamefont {M.~N.}\ \bibnamefont {Armenise}},\
  }\bibfield  {title} {\bibinfo {title} {Recent advances in miniaturized
  optical gyroscopes},\ }\href {https://doi.org/10.2971/jeos.2014.14013}
  {\bibfield  {journal} {\bibinfo  {journal} {Journal of the European Optical
  Society-rapid Publications}\ }\textbf {\bibinfo {volume} {9}},\ \bibinfo
  {pages} {14013} (\bibinfo {year} {2014})}\BibitemShut {NoStop}%
\bibitem [{\citenamefont {Ma}\ \emph {et~al.}(2017)\citenamefont {Ma},
  \citenamefont {Zhang}, \citenamefont {Wang},\ and\ \citenamefont
  {Jin}}]{Ma2017}%
  \BibitemOpen
  \bibfield  {author} {\bibinfo {author} {\bibfnamefont {H.~L.}\ \bibnamefont
  {Ma}}, \bibinfo {author} {\bibfnamefont {J.~J.}\ \bibnamefont {Zhang}},
  \bibinfo {author} {\bibfnamefont {L.~L.}\ \bibnamefont {Wang}},\ and\
  \bibinfo {author} {\bibfnamefont {Z.~H.}\ \bibnamefont {Jin}},\ }\bibfield
  {title} {\bibinfo {title} {Development and evaluation of optical passive
  resonant gyroscopes},\ }\href {https://doi.org/10.1109/JLT.2016.2587667}
  {\bibfield  {journal} {\bibinfo  {journal} {Journal of Lightwave Technology}\
  }\textbf {\bibinfo {volume} {35}},\ \bibinfo {pages} {3546} (\bibinfo {year}
  {2017})}\BibitemShut {NoStop}%
\bibitem [{\citenamefont {Post}(1967)}]{Post1967}%
  \BibitemOpen
  \bibfield  {author} {\bibinfo {author} {\bibfnamefont {E.~J.}\ \bibnamefont
  {Post}},\ }\bibfield  {title} {\bibinfo {title} {Sagnac effect},\ }\href
  {https://doi.org/10.1103/RevModPhys.39.475} {\bibfield  {journal} {\bibinfo
  {journal} {Reviews of Modern Physics}\ }\textbf {\bibinfo {volume} {39}},\
  \bibinfo {pages} {475} (\bibinfo {year} {1967})}\BibitemShut {NoStop}%
\bibitem [{\citenamefont {Ma}\ \emph {et~al.}(2015)\citenamefont {Ma},
  \citenamefont {Zhang}, \citenamefont {Wang},\ and\ \citenamefont
  {Jin}}]{ma2015double}%
  \BibitemOpen
  \bibfield  {author} {\bibinfo {author} {\bibfnamefont {H.}~\bibnamefont
  {Ma}}, \bibinfo {author} {\bibfnamefont {J.}~\bibnamefont {Zhang}}, \bibinfo
  {author} {\bibfnamefont {L.}~\bibnamefont {Wang}},\ and\ \bibinfo {author}
  {\bibfnamefont {Z.}~\bibnamefont {Jin}},\ }\bibfield  {title} {\bibinfo
  {title} {Double closed-loop resonant micro optic gyro using hybrid digital
  phase modulation},\ }\href@noop {} {\bibfield  {journal} {\bibinfo  {journal}
  {Optics express}\ }\textbf {\bibinfo {volume} {23}},\ \bibinfo {pages}
  {15088} (\bibinfo {year} {2015})}\BibitemShut {NoStop}%
\bibitem [{\citenamefont {Wang}\ \emph {et~al.}(2015)\citenamefont {Wang},
  \citenamefont {Feng}, \citenamefont {Tang},\ and\ \citenamefont
  {Zhi}}]{wang2015resonator}%
  \BibitemOpen
  \bibfield  {author} {\bibinfo {author} {\bibfnamefont {J.}~\bibnamefont
  {Wang}}, \bibinfo {author} {\bibfnamefont {L.}~\bibnamefont {Feng}}, \bibinfo
  {author} {\bibfnamefont {Y.}~\bibnamefont {Tang}},\ and\ \bibinfo {author}
  {\bibfnamefont {Y.}~\bibnamefont {Zhi}},\ }\bibfield  {title} {\bibinfo
  {title} {Resonator integrated optic gyro employing trapezoidal phase
  modulation technique},\ }\href@noop {} {\bibfield  {journal} {\bibinfo
  {journal} {Optics letters}\ }\textbf {\bibinfo {volume} {40}},\ \bibinfo
  {pages} {155} (\bibinfo {year} {2015})}\BibitemShut {NoStop}%
\bibitem [{\citenamefont {Liang}\ \emph {et~al.}(2017)\citenamefont {Liang},
  \citenamefont {Ilchenko}, \citenamefont {Savchenkov}, \citenamefont {Dale},
  \citenamefont {Eliyahu}, \citenamefont {Matsko},\ and\ \citenamefont
  {Maleki}}]{liang2017resonant}%
  \BibitemOpen
  \bibfield  {author} {\bibinfo {author} {\bibfnamefont {W.}~\bibnamefont
  {Liang}}, \bibinfo {author} {\bibfnamefont {V.~S.}\ \bibnamefont {Ilchenko}},
  \bibinfo {author} {\bibfnamefont {A.~A.}\ \bibnamefont {Savchenkov}},
  \bibinfo {author} {\bibfnamefont {E.}~\bibnamefont {Dale}}, \bibinfo {author}
  {\bibfnamefont {D.}~\bibnamefont {Eliyahu}}, \bibinfo {author} {\bibfnamefont
  {A.~B.}\ \bibnamefont {Matsko}},\ and\ \bibinfo {author} {\bibfnamefont
  {L.}~\bibnamefont {Maleki}},\ }\bibfield  {title} {\bibinfo {title} {Resonant
  microphotonic gyroscope},\ }\href@noop {} {\bibfield  {journal} {\bibinfo
  {journal} {Optica}\ }\textbf {\bibinfo {volume} {4}},\ \bibinfo {pages} {114}
  (\bibinfo {year} {2017})}\BibitemShut {NoStop}%
\bibitem [{\citenamefont {Li}\ \emph {et~al.}(2017)\citenamefont {Li},
  \citenamefont {Suh},\ and\ \citenamefont {Vahala}}]{Li2017}%
  \BibitemOpen
  \bibfield  {author} {\bibinfo {author} {\bibfnamefont {J.}~\bibnamefont
  {Li}}, \bibinfo {author} {\bibfnamefont {M.-G.}\ \bibnamefont {Suh}},\ and\
  \bibinfo {author} {\bibfnamefont {K.}~\bibnamefont {Vahala}},\ }\bibfield
  {title} {\bibinfo {title} {Microresonator {B}rillouin gyroscope},\ }\href
  {https://doi.org/10.1364/OPTICA.4.000346} {\bibfield  {journal} {\bibinfo
  {journal} {Optica}\ }\textbf {\bibinfo {volume} {4}},\ \bibinfo {pages} {346}
  (\bibinfo {year} {2017})}\BibitemShut {NoStop}%
\bibitem [{\citenamefont {Lai}\ \emph {et~al.}(2019)\citenamefont {Lai},
  \citenamefont {Lu}, \citenamefont {Suh}, \citenamefont {Yuan},\ and\
  \citenamefont {Vahala}}]{lai2019observation}%
  \BibitemOpen
  \bibfield  {author} {\bibinfo {author} {\bibfnamefont {Y.-H.}\ \bibnamefont
  {Lai}}, \bibinfo {author} {\bibfnamefont {Y.-K.}\ \bibnamefont {Lu}},
  \bibinfo {author} {\bibfnamefont {M.-G.}\ \bibnamefont {Suh}}, \bibinfo
  {author} {\bibfnamefont {Z.}~\bibnamefont {Yuan}},\ and\ \bibinfo {author}
  {\bibfnamefont {K.}~\bibnamefont {Vahala}},\ }\bibfield  {title} {\bibinfo
  {title} {Observation of the exceptional-point-enhanced {S}agnac effect},\
  }\href@noop {} {\bibfield  {journal} {\bibinfo  {journal} {Nature}\ }\textbf
  {\bibinfo {volume} {576}},\ \bibinfo {pages} {65} (\bibinfo {year}
  {2019})}\BibitemShut {NoStop}%
\bibitem [{\citenamefont {Wang}\ \emph {et~al.}(2016)\citenamefont {Wang},
  \citenamefont {Feng}, \citenamefont {Wang}, \citenamefont {Jiao},\ and\
  \citenamefont {Wang}}]{Wang2016}%
  \BibitemOpen
  \bibfield  {author} {\bibinfo {author} {\bibfnamefont {J.~J.}\ \bibnamefont
  {Wang}}, \bibinfo {author} {\bibfnamefont {L.~S.}\ \bibnamefont {Feng}},
  \bibinfo {author} {\bibfnamefont {Q.~W.}\ \bibnamefont {Wang}}, \bibinfo
  {author} {\bibfnamefont {H.~C.}\ \bibnamefont {Jiao}},\ and\ \bibinfo
  {author} {\bibfnamefont {X.}~\bibnamefont {Wang}},\ }\bibfield  {title}
  {\bibinfo {title} {Suppression of backreflection error in resonator
  integrated optic gyro by the phase difference traversal method},\ }\href
  {https://doi.org/10.1364/OL.41.001586} {\bibfield  {journal} {\bibinfo
  {journal} {Optics Letters}\ }\textbf {\bibinfo {volume} {41}},\ \bibinfo
  {pages} {1586} (\bibinfo {year} {2016})}\BibitemShut {NoStop}%
\bibitem [{\citenamefont {Khial}\ \emph {et~al.}(2018)\citenamefont {Khial},
  \citenamefont {White},\ and\ \citenamefont {Hajimiri}}]{Khial2018}%
  \BibitemOpen
  \bibfield  {author} {\bibinfo {author} {\bibfnamefont {P.~P.}\ \bibnamefont
  {Khial}}, \bibinfo {author} {\bibfnamefont {A.~D.}\ \bibnamefont {White}},\
  and\ \bibinfo {author} {\bibfnamefont {A.}~\bibnamefont {Hajimiri}},\
  }\bibfield  {title} {\bibinfo {title} {Nanophotonic optical gyroscope with
  reciprocal sensitivity enhancement},\ }\href
  {https://doi.org/10.1038/s41566-018-0266-5} {\bibfield  {journal} {\bibinfo
  {journal} {Nature Photonics}\ }\textbf {\bibinfo {volume} {12}},\ \bibinfo
  {pages} {671} (\bibinfo {year} {2018})}\BibitemShut {NoStop}%
\bibitem [{\citenamefont {Zhang}\ \emph {et~al.}(2016)\citenamefont {Zhang},
  \citenamefont {Liu}, \citenamefont {Lin}, \citenamefont {Li}, \citenamefont
  {Xue}, \citenamefont {Huang},\ and\ \citenamefont {Xiao}}]{Zhang2016a}%
  \BibitemOpen
  \bibfield  {author} {\bibinfo {author} {\bibfnamefont {H.}~\bibnamefont
  {Zhang}}, \bibinfo {author} {\bibfnamefont {J.~M.}\ \bibnamefont {Liu}},
  \bibinfo {author} {\bibfnamefont {J.}~\bibnamefont {Lin}}, \bibinfo {author}
  {\bibfnamefont {W.~X.}\ \bibnamefont {Li}}, \bibinfo {author} {\bibfnamefont
  {X.}~\bibnamefont {Xue}}, \bibinfo {author} {\bibfnamefont {A.~P.}\
  \bibnamefont {Huang}},\ and\ \bibinfo {author} {\bibfnamefont {Z.~S.}\
  \bibnamefont {Xiao}},\ }\bibfield  {title} {\bibinfo {title} {On-chip tunable
  dispersion in a ring laser gyroscope for enhanced rotation sensing},\ }\href
  {https://doi.org/10.1007/s00339-016-0008-9} {\bibfield  {journal} {\bibinfo
  {journal} {Applied Physics A-materials Science \& Processing}\ }\textbf
  {\bibinfo {volume} {122}},\ \bibinfo {pages} {501} (\bibinfo {year}
  {2016})}\BibitemShut {NoStop}%
\bibitem [{\citenamefont {Ren}\ \emph {et~al.}(2017)\citenamefont {Ren},
  \citenamefont {Hodaei}, \citenamefont {Harari}, \citenamefont {Hassan},
  \citenamefont {Chow}, \citenamefont {Soltani}, \citenamefont
  {Christodoulides},\ and\ \citenamefont {Khajavikhan}}]{Ren2017}%
  \BibitemOpen
  \bibfield  {author} {\bibinfo {author} {\bibfnamefont {J.}~\bibnamefont
  {Ren}}, \bibinfo {author} {\bibfnamefont {H.}~\bibnamefont {Hodaei}},
  \bibinfo {author} {\bibfnamefont {G.}~\bibnamefont {Harari}}, \bibinfo
  {author} {\bibfnamefont {A.~U.}\ \bibnamefont {Hassan}}, \bibinfo {author}
  {\bibfnamefont {W.}~\bibnamefont {Chow}}, \bibinfo {author} {\bibfnamefont
  {M.}~\bibnamefont {Soltani}}, \bibinfo {author} {\bibfnamefont
  {D.}~\bibnamefont {Christodoulides}},\ and\ \bibinfo {author} {\bibfnamefont
  {M.}~\bibnamefont {Khajavikhan}},\ }\bibfield  {title} {\bibinfo {title}
  {Ultrasensitive micro-scale parity-time-symmetric ring laser gyroscope},\
  }\href {https://doi.org/10.1364/OL.42.001556} {\bibfield  {journal} {\bibinfo
   {journal} {Optics Letters}\ }\textbf {\bibinfo {volume} {42}},\ \bibinfo
  {pages} {1556} (\bibinfo {year} {2017})}\BibitemShut {NoStop}%
\bibitem [{\citenamefont {Sunada}(2017)}]{Sunada2017}%
  \BibitemOpen
  \bibfield  {author} {\bibinfo {author} {\bibfnamefont {S.}~\bibnamefont
  {Sunada}},\ }\bibfield  {title} {\bibinfo {title} {Large {S}agnac frequency
  splitting in a ring resonator operating at an exceptional point},\ }\href
  {https://doi.org/10.1103/PhysRevA.96.033842} {\bibfield  {journal} {\bibinfo
  {journal} {Physical Review A}\ }\textbf {\bibinfo {volume} {96}},\ \bibinfo
  {pages} {033842} (\bibinfo {year} {2017})}\BibitemShut {NoStop}%
\bibitem [{\citenamefont {Del'Haye}\ \emph {et~al.}(2007)\citenamefont
  {Del'Haye}, \citenamefont {Schliesser}, \citenamefont {Arcizet},
  \citenamefont {Wilken}, \citenamefont {Holzwarth},\ and\ \citenamefont
  {Kippenberg}}]{Del'Haye2007}%
  \BibitemOpen
  \bibfield  {author} {\bibinfo {author} {\bibfnamefont {P.}~\bibnamefont
  {Del'Haye}}, \bibinfo {author} {\bibfnamefont {A.}~\bibnamefont
  {Schliesser}}, \bibinfo {author} {\bibfnamefont {O.}~\bibnamefont {Arcizet}},
  \bibinfo {author} {\bibfnamefont {T.}~\bibnamefont {Wilken}}, \bibinfo
  {author} {\bibfnamefont {R.}~\bibnamefont {Holzwarth}},\ and\ \bibinfo
  {author} {\bibfnamefont {T.~J.}\ \bibnamefont {Kippenberg}},\ }\bibfield
  {title} {\bibinfo {title} {Optical frequency comb generation from a
  monolithic microresonator},\ }\href@noop {} {\bibfield  {journal} {\bibinfo
  {journal} {Nature}\ }\textbf {\bibinfo {volume} {450}},\ \bibinfo {pages}
  {1214} (\bibinfo {year} {2007})}\BibitemShut {NoStop}%
\bibitem [{\citenamefont {Stern}\ \emph {et~al.}(2018)\citenamefont {Stern},
  \citenamefont {Ji}, \citenamefont {Okawachi}, \citenamefont {Gaeta},\ and\
  \citenamefont {Lipson}}]{Stern2018}%
  \BibitemOpen
  \bibfield  {author} {\bibinfo {author} {\bibfnamefont {B.}~\bibnamefont
  {Stern}}, \bibinfo {author} {\bibfnamefont {X.~C.}\ \bibnamefont {Ji}},
  \bibinfo {author} {\bibfnamefont {Y.}~\bibnamefont {Okawachi}}, \bibinfo
  {author} {\bibfnamefont {A.~L.}\ \bibnamefont {Gaeta}},\ and\ \bibinfo
  {author} {\bibfnamefont {M.}~\bibnamefont {Lipson}},\ }\bibfield  {title}
  {\bibinfo {title} {Battery-operated integrated frequency comb generator},\
  }\href {https://doi.org/10.1038/s41586-018-0598-9} {\bibfield  {journal}
  {\bibinfo  {journal} {Nature}\ }\textbf {\bibinfo {volume} {562}},\ \bibinfo
  {pages} {401} (\bibinfo {year} {2018})}\BibitemShut {NoStop}%
\bibitem [{\citenamefont {Zhang}\ \emph {et~al.}(2019)\citenamefont {Zhang},
  \citenamefont {Silver}, \citenamefont {Bino}, \citenamefont {Copie},
  \citenamefont {Woodley}, \citenamefont {Ghalanos}, \citenamefont {Svela},
  \citenamefont {Moroney},\ and\ \citenamefont {Del'Haye}}]{zhang2019sub}%
  \BibitemOpen
  \bibfield  {author} {\bibinfo {author} {\bibfnamefont {S.}~\bibnamefont
  {Zhang}}, \bibinfo {author} {\bibfnamefont {J.~M.}\ \bibnamefont {Silver}},
  \bibinfo {author} {\bibfnamefont {L.~D.}\ \bibnamefont {Bino}}, \bibinfo
  {author} {\bibfnamefont {F.}~\bibnamefont {Copie}}, \bibinfo {author}
  {\bibfnamefont {M.~T.~M.}\ \bibnamefont {Woodley}}, \bibinfo {author}
  {\bibfnamefont {G.~N.}\ \bibnamefont {Ghalanos}}, \bibinfo {author}
  {\bibfnamefont {A.~{\O}.}\ \bibnamefont {Svela}}, \bibinfo {author}
  {\bibfnamefont {N.}~\bibnamefont {Moroney}},\ and\ \bibinfo {author}
  {\bibfnamefont {P.}~\bibnamefont {Del'Haye}},\ }\bibfield  {title} {\bibinfo
  {title} {Sub-milliwatt-level microresonator solitons with extended access
  range using an auxiliary laser},\ }\href
  {https://doi.org/10.1364/OPTICA.6.000206} {\bibfield  {journal} {\bibinfo
  {journal} {Optica}\ }\textbf {\bibinfo {volume} {6}},\ \bibinfo {pages} {206}
  (\bibinfo {year} {2019})}\BibitemShut {NoStop}%
\bibitem [{\citenamefont {Del~Bino}\ \emph {et~al.}(2017)\citenamefont
  {Del~Bino}, \citenamefont {Silver}, \citenamefont {Stebbings},\ and\
  \citenamefont {Del'Haye}}]{del2017symmetry}%
  \BibitemOpen
  \bibfield  {author} {\bibinfo {author} {\bibfnamefont {L.}~\bibnamefont
  {Del~Bino}}, \bibinfo {author} {\bibfnamefont {J.~M.}\ \bibnamefont
  {Silver}}, \bibinfo {author} {\bibfnamefont {S.~L.}\ \bibnamefont
  {Stebbings}},\ and\ \bibinfo {author} {\bibfnamefont {P.}~\bibnamefont
  {Del'Haye}},\ }\bibfield  {title} {\bibinfo {title} {Symmetry breaking of
  counter-propagating light in a nonlinear resonator},\ }\href@noop {}
  {\bibfield  {journal} {\bibinfo  {journal} {Scientific Reports}\ }\textbf
  {\bibinfo {volume} {7}} (\bibinfo {year} {2017})}\BibitemShut {NoStop}%
\bibitem [{\citenamefont {Cao}\ \emph {et~al.}(2017)\citenamefont {Cao},
  \citenamefont {Wang}, \citenamefont {Dong}, \citenamefont {Jing},
  \citenamefont {Liu}, \citenamefont {Chen}, \citenamefont {Ge}, \citenamefont
  {Gong},\ and\ \citenamefont {Xiao}}]{Cao2017}%
  \BibitemOpen
  \bibfield  {author} {\bibinfo {author} {\bibfnamefont {Q.~T.}\ \bibnamefont
  {Cao}}, \bibinfo {author} {\bibfnamefont {H.~M.}\ \bibnamefont {Wang}},
  \bibinfo {author} {\bibfnamefont {C.~H.}\ \bibnamefont {Dong}}, \bibinfo
  {author} {\bibfnamefont {H.}~\bibnamefont {Jing}}, \bibinfo {author}
  {\bibfnamefont {R.~S.}\ \bibnamefont {Liu}}, \bibinfo {author} {\bibfnamefont
  {X.}~\bibnamefont {Chen}}, \bibinfo {author} {\bibfnamefont {L.}~\bibnamefont
  {Ge}}, \bibinfo {author} {\bibfnamefont {Q.~H.}\ \bibnamefont {Gong}},\ and\
  \bibinfo {author} {\bibfnamefont {Y.~F.}\ \bibnamefont {Xiao}},\ }\bibfield
  {title} {\bibinfo {title} {Experimental demonstration of spontaneous
  chirality in a nonlinear microresonator},\ }\href
  {https://doi.org/10.1103/PhysRevLett.118.033901} {\bibfield  {journal}
  {\bibinfo  {journal} {Physical Review Letters}\ }\textbf {\bibinfo {volume}
  {118}},\ \bibinfo {pages} {033901} (\bibinfo {year} {2017})}\BibitemShut
  {NoStop}%
\bibitem [{\citenamefont {Woodley}\ \emph {et~al.}(2018)\citenamefont
  {Woodley}, \citenamefont {Silver}, \citenamefont {Hill}, \citenamefont
  {Copie}, \citenamefont {Del~Bino}, \citenamefont {Zhang}, \citenamefont
  {Oppo},\ and\ \citenamefont {Del'Haye}}]{Woodley2018}%
  \BibitemOpen
  \bibfield  {author} {\bibinfo {author} {\bibfnamefont {M.~T.~M.}\
  \bibnamefont {Woodley}}, \bibinfo {author} {\bibfnamefont {J.~M.}\
  \bibnamefont {Silver}}, \bibinfo {author} {\bibfnamefont {L.}~\bibnamefont
  {Hill}}, \bibinfo {author} {\bibfnamefont {F.}~\bibnamefont {Copie}},
  \bibinfo {author} {\bibfnamefont {L.}~\bibnamefont {Del~Bino}}, \bibinfo
  {author} {\bibfnamefont {S.~Y.}\ \bibnamefont {Zhang}}, \bibinfo {author}
  {\bibfnamefont {G.~L.}\ \bibnamefont {Oppo}},\ and\ \bibinfo {author}
  {\bibfnamefont {P.}~\bibnamefont {Del'Haye}},\ }\bibfield  {title} {\bibinfo
  {title} {Universal symmetry-breaking dynamics for the {K}err interaction of
  counterpropagating light in dielectric ring resonators},\ }\href
  {https://doi.org/10.1103/PhysRevA.98.053863} {\bibfield  {journal} {\bibinfo
  {journal} {Physical Review A}\ }\textbf {\bibinfo {volume} {98}},\ \bibinfo
  {pages} {053863} (\bibinfo {year} {2018})}\BibitemShut {NoStop}%
\bibitem [{\citenamefont {Del~Bino}\ \emph {et~al.}(2018)\citenamefont
  {Del~Bino}, \citenamefont {Silver}, \citenamefont {Woodley}, \citenamefont
  {Stebbings}, \citenamefont {Zhao},\ and\ \citenamefont
  {Del'Haye}}]{DelBino2018}%
  \BibitemOpen
  \bibfield  {author} {\bibinfo {author} {\bibfnamefont {L.}~\bibnamefont
  {Del~Bino}}, \bibinfo {author} {\bibfnamefont {J.~M.}\ \bibnamefont
  {Silver}}, \bibinfo {author} {\bibfnamefont {M.~T.~M.}\ \bibnamefont
  {Woodley}}, \bibinfo {author} {\bibfnamefont {S.~L.}\ \bibnamefont
  {Stebbings}}, \bibinfo {author} {\bibfnamefont {X.}~\bibnamefont {Zhao}},\
  and\ \bibinfo {author} {\bibfnamefont {P.}~\bibnamefont {Del'Haye}},\
  }\bibfield  {title} {\bibinfo {title} {Microresonator isolators and
  circulators based on the intrinsic nonreciprocity of the {K}err effect},\
  }\href {https://doi.org/10.1364/OPTICA.5.000279} {\bibfield  {journal}
  {\bibinfo  {journal} {Optica}\ }\textbf {\bibinfo {volume} {5}},\ \bibinfo
  {pages} {279} (\bibinfo {year} {2018})}\BibitemShut {NoStop}%
\bibitem [{\citenamefont {Daniel}\ and\ \citenamefont
  {Agrawal}(2012)}]{Daniel2012}%
  \BibitemOpen
  \bibfield  {author} {\bibinfo {author} {\bibfnamefont {B.~A.}\ \bibnamefont
  {Daniel}}\ and\ \bibinfo {author} {\bibfnamefont {G.~P.}\ \bibnamefont
  {Agrawal}},\ }\bibfield  {title} {\bibinfo {title} {Phase-switched
  all-optical flip-flops using two-input bistable resonators},\ }\href
  {https://doi.org/10.1109/LPT.2011.2181832} {\bibfield  {journal} {\bibinfo
  {journal} {IEEE Photonics Technology Letters}\ }\textbf {\bibinfo {volume}
  {24}},\ \bibinfo {pages} {479} (\bibinfo {year} {2012})}\BibitemShut
  {NoStop}%
\bibitem [{\citenamefont {Kaplan}\ and\ \citenamefont
  {Meystre}(1981)}]{Kaplan1981}%
  \BibitemOpen
  \bibfield  {author} {\bibinfo {author} {\bibfnamefont {A.~E.}\ \bibnamefont
  {Kaplan}}\ and\ \bibinfo {author} {\bibfnamefont {P.}~\bibnamefont
  {Meystre}},\ }\bibfield  {title} {\bibinfo {title} {Enhancement of the
  {S}agnac effect due to nonlinearly induced nonreciprocity},\ }\href
  {https://doi.org/10.1364/OL.6.000590} {\bibfield  {journal} {\bibinfo
  {journal} {Optics Letters}\ }\textbf {\bibinfo {volume} {6}},\ \bibinfo
  {pages} {590} (\bibinfo {year} {1981})}\BibitemShut {NoStop}%
\bibitem [{\citenamefont {Wang}\ and\ \citenamefont
  {Search}(2014)}]{Wang2014a}%
  \BibitemOpen
  \bibfield  {author} {\bibinfo {author} {\bibfnamefont {C.}~\bibnamefont
  {Wang}}\ and\ \bibinfo {author} {\bibfnamefont {C.~P.}\ \bibnamefont
  {Search}},\ }\bibfield  {title} {\bibinfo {title} {Enhanced rotation sensing
  by nonlinear interactions in silicon microresonators},\ }\href
  {https://doi.org/10.1364/OL.39.004376} {\bibfield  {journal} {\bibinfo
  {journal} {Optics Letters}\ }\textbf {\bibinfo {volume} {39}},\ \bibinfo
  {pages} {4376} (\bibinfo {year} {2014})}\BibitemShut {NoStop}%
\bibitem [{\citenamefont {Wang}\ and\ \citenamefont {Search}(2015)}]{Wang2015}%
  \BibitemOpen
  \bibfield  {author} {\bibinfo {author} {\bibfnamefont {C.}~\bibnamefont
  {Wang}}\ and\ \bibinfo {author} {\bibfnamefont {C.~P.}\ \bibnamefont
  {Search}},\ }\bibfield  {title} {\bibinfo {title} {A nonlinear microresonator
  refractive index sensor},\ }\href {https://doi.org/10.1109/JLT.2015.2464105}
  {\bibfield  {journal} {\bibinfo  {journal} {Journal of Lightwave Technology}\
  }\textbf {\bibinfo {volume} {33}},\ \bibinfo {pages} {4360} (\bibinfo {year}
  {2015})}\BibitemShut {NoStop}%
\bibitem [{\citenamefont {Svela}\ \emph {et~al.}(2019)\citenamefont {Svela},
  \citenamefont {Silver}, \citenamefont {Del~Bino}, \citenamefont {Ghalanos},
  \citenamefont {Moroney}, \citenamefont {Woodley}, \citenamefont {Zhang},
  \citenamefont {Vanner},\ and\ \citenamefont
  {Del'Haye}}]{svela2019spontaneous}%
  \BibitemOpen
  \bibfield  {author} {\bibinfo {author} {\bibfnamefont {A.~{\O}.}\
  \bibnamefont {Svela}}, \bibinfo {author} {\bibfnamefont {J.~M.}\ \bibnamefont
  {Silver}}, \bibinfo {author} {\bibfnamefont {L.}~\bibnamefont {Del~Bino}},
  \bibinfo {author} {\bibfnamefont {G.}~\bibnamefont {Ghalanos}}, \bibinfo
  {author} {\bibfnamefont {N.}~\bibnamefont {Moroney}}, \bibinfo {author}
  {\bibfnamefont {M.~T.}\ \bibnamefont {Woodley}}, \bibinfo {author}
  {\bibfnamefont {S.}~\bibnamefont {Zhang}}, \bibinfo {author} {\bibfnamefont
  {M.}~\bibnamefont {Vanner}},\ and\ \bibinfo {author} {\bibfnamefont
  {P.}~\bibnamefont {Del'Haye}},\ }\bibfield  {title} {\bibinfo {title}
  {Spontaneous symmetry breaking based near-field sensing with a
  microresonator},\ }in\ \href@noop {} {\emph {\bibinfo {booktitle} {CLEO:
  QELS\_Fundamental Science}}}\ (\bibinfo {organization} {Optical Society of
  America},\ \bibinfo {year} {2019})\ pp.\ \bibinfo {pages}
  {JM3B--3}\BibitemShut {NoStop}%
\bibitem [{\citenamefont {Stanley}(1971)}]{stanley1971phase}%
  \BibitemOpen
  \bibfield  {author} {\bibinfo {author} {\bibfnamefont {H.~E.}\ \bibnamefont
  {Stanley}},\ }\href@noop {} {\emph {\bibinfo {title} {Phase transitions and
  critical phenomena}}}\ (\bibinfo  {publisher} {Clarendon Press, Oxford},\
  \bibinfo {year} {1971})\BibitemShut {NoStop}%
\bibitem [{\citenamefont {Higgs}(1964)}]{higgs1964broken}%
  \BibitemOpen
  \bibfield  {author} {\bibinfo {author} {\bibfnamefont {P.~W.}\ \bibnamefont
  {Higgs}},\ }\bibfield  {title} {\bibinfo {title} {Broken symmetries and the
  masses of gauge bosons},\ }\href@noop {} {\bibfield  {journal} {\bibinfo
  {journal} {Physical Review Letters}\ }\textbf {\bibinfo {volume} {13}},\
  \bibinfo {pages} {508} (\bibinfo {year} {1964})}\BibitemShut {NoStop}%
\bibitem [{\citenamefont {Bardeen}\ \emph {et~al.}(1957)\citenamefont
  {Bardeen}, \citenamefont {Cooper},\ and\ \citenamefont
  {Schrieffer}}]{PhysRev.108.1175}%
  \BibitemOpen
  \bibfield  {author} {\bibinfo {author} {\bibfnamefont {J.}~\bibnamefont
  {Bardeen}}, \bibinfo {author} {\bibfnamefont {L.~N.}\ \bibnamefont
  {Cooper}},\ and\ \bibinfo {author} {\bibfnamefont {J.~R.}\ \bibnamefont
  {Schrieffer}},\ }\bibfield  {title} {\bibinfo {title} {Theory of
  superconductivity},\ }\href {https://doi.org/10.1103/PhysRev.108.1175}
  {\bibfield  {journal} {\bibinfo  {journal} {Phys. Rev.}\ }\textbf {\bibinfo
  {volume} {108}},\ \bibinfo {pages} {1175} (\bibinfo {year}
  {1957})}\BibitemShut {NoStop}%
\bibitem [{\citenamefont {Landau}(1941)}]{landau1941theory}%
  \BibitemOpen
  \bibfield  {author} {\bibinfo {author} {\bibfnamefont {L.}~\bibnamefont
  {Landau}},\ }\bibfield  {title} {\bibinfo {title} {Theory of the
  superfluidity of helium {II}},\ }\href@noop {} {\bibfield  {journal}
  {\bibinfo  {journal} {Physical Review}\ }\textbf {\bibinfo {volume} {60}},\
  \bibinfo {pages} {356} (\bibinfo {year} {1941})}\BibitemShut {NoStop}%
\bibitem [{\citenamefont {Silver}\ \emph {et~al.}(2019)\citenamefont {Silver},
  \citenamefont {Grattan},\ and\ \citenamefont
  {Del'Haye}}]{silver2019criticalPaperArxiv}%
  \BibitemOpen
  \bibfield  {author} {\bibinfo {author} {\bibfnamefont {J.~M.}\ \bibnamefont
  {Silver}}, \bibinfo {author} {\bibfnamefont {K.~T.}\ \bibnamefont
  {Grattan}},\ and\ \bibinfo {author} {\bibfnamefont {P.}~\bibnamefont
  {Del'Haye}},\ }\bibfield  {title} {\bibinfo {title} {Critical dynamics of an
  asymmetrically bidirectionally pumped optical microresonator},\ }\href@noop
  {} {\bibfield  {journal} {\bibinfo  {journal} {arXiv preprint
  arXiv:1912.08262}\ } (\bibinfo {year} {2019})}\BibitemShut {NoStop}%
\bibitem [{\citenamefont {Del'Haye}\ \emph {et~al.}(2013)\citenamefont
  {Del'Haye}, \citenamefont {Diddams},\ and\ \citenamefont
  {Papp}}]{DelHaye2013apl}%
  \BibitemOpen
  \bibfield  {author} {\bibinfo {author} {\bibfnamefont {P.}~\bibnamefont
  {Del'Haye}}, \bibinfo {author} {\bibfnamefont {S.~A.}\ \bibnamefont
  {Diddams}},\ and\ \bibinfo {author} {\bibfnamefont {S.~B.}\ \bibnamefont
  {Papp}},\ }\bibfield  {title} {\bibinfo {title} {Laser-machined
  ultra-high-{Q} microrod resonators for nonlinear optics},\ }\href@noop {}
  {\bibfield  {journal} {\bibinfo  {journal} {Applied Physics Letters}\
  }\textbf {\bibinfo {volume} {102}},\ \bibinfo {pages} {221119} (\bibinfo
  {year} {2013})}\BibitemShut {NoStop}%
\bibitem [{\citenamefont {Lee}\ \emph {et~al.}(2012)\citenamefont {Lee},
  \citenamefont {Chen}, \citenamefont {Li}, \citenamefont {Yang}, \citenamefont
  {Jeon}, \citenamefont {Painter},\ and\ \citenamefont {Vahala}}]{Lee2012}%
  \BibitemOpen
  \bibfield  {author} {\bibinfo {author} {\bibfnamefont {H.}~\bibnamefont
  {Lee}}, \bibinfo {author} {\bibfnamefont {T.}~\bibnamefont {Chen}}, \bibinfo
  {author} {\bibfnamefont {J.}~\bibnamefont {Li}}, \bibinfo {author}
  {\bibfnamefont {K.~Y.}\ \bibnamefont {Yang}}, \bibinfo {author}
  {\bibfnamefont {S.}~\bibnamefont {Jeon}}, \bibinfo {author} {\bibfnamefont
  {O.}~\bibnamefont {Painter}},\ and\ \bibinfo {author} {\bibfnamefont {K.~J.}\
  \bibnamefont {Vahala}},\ }\bibfield  {title} {\bibinfo {title} {Chemically
  etched ultrahigh-{Q} wedge-resonator on a silicon chip},\ }\href
  {https://doi.org/10.1038/NPHOTON.2012.109} {\bibfield  {journal} {\bibinfo
  {journal} {Nature Photonics}\ }\textbf {\bibinfo {volume} {6}},\ \bibinfo
  {pages} {369} (\bibinfo {year} {2012})}\BibitemShut {NoStop}%
\bibitem [{\citenamefont {Passaro}\ \emph {et~al.}(2017)\citenamefont
  {Passaro}, \citenamefont {Cuccovillo}, \citenamefont {Vaiani}, \citenamefont
  {De~Carlo},\ and\ \citenamefont {Campanella}}]{Passaro2017}%
  \BibitemOpen
  \bibfield  {author} {\bibinfo {author} {\bibfnamefont {V.~M.~N.}\
  \bibnamefont {Passaro}}, \bibinfo {author} {\bibfnamefont {A.}~\bibnamefont
  {Cuccovillo}}, \bibinfo {author} {\bibfnamefont {L.}~\bibnamefont {Vaiani}},
  \bibinfo {author} {\bibfnamefont {M.}~\bibnamefont {De~Carlo}},\ and\
  \bibinfo {author} {\bibfnamefont {C.~E.}\ \bibnamefont {Campanella}},\
  }\bibfield  {title} {\bibinfo {title} {Gyroscope technology and applications:
  A review in the industrial perspective},\ }\href
  {https://doi.org/10.3390/s17102284} {\bibfield  {journal} {\bibinfo
  {journal} {Sensors}\ }\textbf {\bibinfo {volume} {17}},\ \bibinfo {pages}
  {2284} (\bibinfo {year} {2017})}\BibitemShut {NoStop}%
\bibitem [{\citenamefont {Lecaplain}\ \emph {et~al.}(2016)\citenamefont
  {Lecaplain}, \citenamefont {Javerzac-Galy}, \citenamefont {Gorodetsky},\ and\
  \citenamefont {Kippenberg}}]{Lecaplain2016}%
  \BibitemOpen
  \bibfield  {author} {\bibinfo {author} {\bibfnamefont {C.}~\bibnamefont
  {Lecaplain}}, \bibinfo {author} {\bibfnamefont {C.}~\bibnamefont
  {Javerzac-Galy}}, \bibinfo {author} {\bibfnamefont {M.~L.}\ \bibnamefont
  {Gorodetsky}},\ and\ \bibinfo {author} {\bibfnamefont {T.~J.}\ \bibnamefont
  {Kippenberg}},\ }\bibfield  {title} {\bibinfo {title} {Mid-infrared
  ultra-high-{Q} resonators based on fluoride crystalline materials},\ }\href
  {https://doi.org/10.1038/ncomms13383} {\bibfield  {journal} {\bibinfo
  {journal} {Nature Communications}\ }\textbf {\bibinfo {volume} {7}},\
  \bibinfo {pages} {13383} (\bibinfo {year} {2016})}\BibitemShut {NoStop}%
\bibitem [{\citenamefont {Carmon}\ \emph {et~al.}(2004)\citenamefont {Carmon},
  \citenamefont {Yang},\ and\ \citenamefont {Vahala}}]{Carmon2004}%
  \BibitemOpen
  \bibfield  {author} {\bibinfo {author} {\bibfnamefont {T.}~\bibnamefont
  {Carmon}}, \bibinfo {author} {\bibfnamefont {L.}~\bibnamefont {Yang}},\ and\
  \bibinfo {author} {\bibfnamefont {K.~J.}\ \bibnamefont {Vahala}},\ }\bibfield
   {title} {\bibinfo {title} {Dynamical thermal behavior and thermal
  self-stability of microcavities},\ }\href@noop {} {\bibfield  {journal}
  {\bibinfo  {journal} {Optics Express}\ }\textbf {\bibinfo {volume} {12}},\
  \bibinfo {pages} {4742} (\bibinfo {year} {2004})}\BibitemShut {NoStop}%
\end{thebibliography}%
\end{document}